\documentclass[twocolumn]{aastex631}

\shorttitle{Rotation in a $z=9.1$ Galaxy}
\shortauthors{Tokuoka et al.}
\graphicspath{{./}{figures/}}

\begin{document}

\title{Possible Systematic Rotation in the Mature Stellar Population of a $z=9.1$ Galaxy}

\author{Tsuyoshi Tokuoka}
\affiliation{Department of Pure and Applied Physics, Graduate School of Advanced Science and Engineering,\\
Faculty of Science and Engineering, Waseda University, 3-4-1, Okubo, Shinjuku, Tokyo 169-8555, Japan}

\author[0000-0002-7779-8677]{Akio K.\ Inoue}
\affiliation{Department of Pure and Applied Physics, Graduate School of Advanced Science and Engineering,\\
Faculty of Science and Engineering, Waseda University, 3-4-1, Okubo, Shinjuku, Tokyo 169-8555, Japan}
\affiliation{Waseda Research Institute for Science and Engineering, Faculty of Science and Engineering, Waseda University,\\
3-4-1, Okubo, Shinjuku, Tokyo 169-8555, Japan}
\email{akinoue@aoni.waseda.jp}

\author[0000-0002-0898-4038]{Takuya Hashimoto}
\affiliation{Tomonaga Center for the History of the Universe (TCHoU), Faculty of Pure and Applied Sciences, University of Tsukuba, Tsukuba, Ibaraki 305-8571, Japan}


\author{Richard S.\ Ellis}
\affiliation{Department of Physics \& Astronomy, University College London, Gower Street, London WC1E 6BT, United Kingdom}

\author[0000-0001-7459-6335]{Nicolas Laporte}
\affiliation{Kavli Institute for Cosmology, University of Cambridge, Madingley Road, Cambridge CB3 0HA, UK} 
\affiliation{Cavendish Laboratory, University of Cambridge, 19 JJ Thomson Avenue, Cambridge CB3 0HE, UK}

\author[0000-0001-6958-7856]{Yuma Sugahara}
\affiliation{Waseda Research Institute for Science and Engineering, Faculty of Science and Engineering, Waseda University,\\
3-4-1, Okubo, Shinjuku, Tokyo 169-8555, Japan}
\affiliation{National Astronomical Observatory of Japan, 2-21-1, Osawa, Mitaka, Tokyo 181-8588, Japan}

\author[0000-0003-3278-2484]{Hiroshi Matsuo}
\affiliation{National Astronomical Observatory of Japan, 2-21-1, Osawa, Mitaka, Tokyo 181-8588, Japan}

\author[0000-0003-4807-8117]{Yoichi Tamura}
\affiliation{Division of Particle and Astrophysical Science, Graduate School of Science, Nagoya University, Aichi 4648602, Japan}


\author{Yoshinobu Fudamoto}
\affiliation{Waseda Research Institute for Science and Engineering, Faculty of Science and Engineering, Waseda University,\\
3-4-1, Okubo, Shinjuku, Tokyo 169-8555, Japan}
\affiliation{National Astronomical Observatory of Japan, 2-21-1, Osawa, Mitaka, Tokyo 181-8588, Japan}

\author[0000-0003-3349-4070]{Kana Moriwaki}
\affiliation{Department of Physics, The University of Tokyo, 7-3-1 Hongo, Bunkyo, Tokyo 113-0033, Japan}

\author[0000-0002-4140-1367]{Guido Roberts-Borsani}
\affiliation{Department of Physics and Astronomy, University of California, Los Angeles, 430 Portola Plaza, Los Angeles, CA 90095, USA}

\author{Ikkoh Shimizu}
\affiliation{Department of Literature, Shikoku Gakuin University, 3-2-1 Bunkyocho, Zentsuji, Kagawa 765-8505, Japan}

\author[0000-0002-7738-5290]{Satoshi Yamanaka}
\affiliation{General Education Department, National Institute of Technology, Toba College, 1-1, Ikegami-cho, Toba, Mie 517-8501, Japan}

\author[0000-0001-7925-238X]{Naoki Yoshida}
\affiliation{Department of Physics, The University of Tokyo, 7-3-1 Hongo, Bunkyo, Tokyo 113-0033, Japan}
\affiliation{Kavli Institute for the Physics and Mathematics of the Universe (WPI), UT Institutes for Advanced Study, The University of Tokyo, 5-1-5 Kashiwanoha, Kashiwa, Chiba 277-8583, Japan}
\affiliation{Research Center for the Early Universe, School of Science, The University of Tokyo, 7-3-1 Hongo, Bunkyo, Tokyo 113-0033, Japan}
\affiliation{Institute for Physics of Intelligence, School of Science, The University of Tokyo, 7-3-1 Hongo, Bunkyo, Tokyo 113-0033, Japan}

\author[0000-0003-1096-2636]{Erik Zackrisson}
\affiliation{Observational Astrophysics, Department of Physics and Astronomy, Uppsala University, Box 516, SE-751 20 Uppsala, Sweden}

\author[0000-0002-0205-5174]{Wei Zheng}
\affiliation{Department of Physics and Astronomy, Johns Hopkins University, Baltimore, MD 21218, USA}



\begin{abstract}
We present new observations with the Atacama Large Millimeter/submillimeter Array for a gravitationally-lensed galaxy at $z=9.1$, MACS1149-JD1.
[O~{\sc iii}] 88-$\mu$m emission is detected at 10$\sigma$ with a spatial resolution of $\sim0.3$ kpc in the source plane, enabling the most distant morpho-kinematic study of a galaxy.
The [O~{\sc iii}] emission is distributed smoothly without any resolved clumps and shows a clear velocity gradient with $\Delta V_{\rm obs}/2\sigma_{\rm tot}=0.84\pm0.23$, where $\Delta V_{\rm obs}$ is the observed maximum velocity difference and $\sigma_{\rm tot}$ is the velocity dispersion measured in the spatially-integrated line profile, suggesting a rotating system.
Assuming a geometrically thin self-gravitating rotation disk model, we obtain $V_{\rm rot}/\sigma_V=0.67_{-0.26}^{+0.73}$, where $V_{\rm rot}$ and $\sigma_V$ are the rotation velocity and velocity dispersion, respectively, still consistent with rotation.
The resulting disk mass of $0.65_{-0.40}^{+1.37}\times10^{9}$ M$_\odot$ is consistent with being associated with the stellar mass identified with a 300 Myr-old stellar population independently indicated by a Balmer break in the spectral energy distribution.
We conclude that the most of the dynamical mass is associated with the previously-identified mature stellar population that formed at $z\sim15$.
\end{abstract}



\section{Introduction} \label{sec:intro}

The Atacama Large Millimeter/submillimeter Array (ALMA) has revolutionized high-redshift galaxy observations, allowing galaxies to be characterized well into the epoch of reionization.
For example, dust continuum as well as the [O~{\sc iii}] 88- and [C~{\sc ii}] 158-$\mu$m emission lines of galaxies at $z>7$ have been successfully observed (e.g., \citealt{2015Natur.519..327W,2016Sci...352.1559I,2019PASJ...71...71H}).
In particular, MACS1149-JD1 \citep{2012Natur.489..406Z,2014ApJ...795..126B,2016ApJ...817...11H,2017ApJ...836..210Z,2018ApJ...854...39H} is a gravitationally-lensed galaxy emitting the [O~{\sc iii}] line at $z=9.1$, one of the most distant objects spectroscopically confirmed (\citealt{2018Natur.557..392H}, hereafter H18).
This galaxy also shows a Balmer break consistent with a stellar population of a few hundred Myrs old, suggesting its formation epoch is $z\sim15$ (H18, \citealt{2019MNRAS.489.3827B,2020MNRAS.497.3440R,2021MNRAS.505.3336L}).

Beyond finding high-redshift galaxies, studying their dynamics based on the kinematics of their interstellar medium provides further motivation for probing the early physics of galaxy formation.
Such studies have been mostly conducted with three-dimensional (3D) near-infrared spectroscopy for galaxies at $z<4$ (e.g., \citealt{2009ApJ...706.1364F,2010MNRAS.404.1247J,2015ApJ...799..209W}).
However, the high sensitivity and high spatial and frequency resolution of ALMA also make it possible to analyze morpho-kinematics of galaxies at $4<z<6$ \citep{2020Natur.584..201R,2021MNRAS.507.3952R,2021Sci...371..713L} and even at $z\sim7$ \citep{2018Natur.553..178S}.
In this {\it Letter}, we present the most distant example of a morpho-kinematic analysis of the [O~{\sc iii}] emission in MACS1149-JD1 at $z=9.1$ and discuss when the rotational motion in galaxies first appears.

Throughout this {\it Letter}, we use a flat $\Lambda$CDM cosmology with a parameter set of $(h,\Omega_m,\Omega_\Lambda)=(0.704,0.272,0.728)$ \citep{2011ApJS..192...18K}.

\section{Observational data}

New observations for [O~{\sc iii}] 88-$\mu$m emission from MACS1149-JD1 were performed in Band 7 during ALMA Cycle 6 (2018.1.00616.S, PI: T.~Hashimoto) to improve the spatial resolution of the emission line.
The antenna configurations were C43-4, -5, and -6 (minimum baseline = 15.1 m and maximum baseline = 783.5--2516.9 m).
We set a spectral window (SPW) centered at a 335.625-GHz emission line frequency (H18) with a bandwidth of 1.875 GHz and 240 channels, corresponding to a velocity resolution of 7.0 km s$^{-1}$.
Three other SPWs were set at the central frequencies of 337.375, 347.417, and 349.176 GHz to obtain the continuum emission, with bandwidths of 2.000 GHz and 128 channels.
The observations were executed in a series of 13 sets between October 18 and December 15, 2018.
The precipitable water vapor during the observations spread over 0.3--1.0 mm, and the mean was 0.5 mm.
The total on-source exposure time was 9.6 h, compared to 2.0 h in the previous observations (H18).
Raw data were processed using Common Astronomy Software Applications (CASA; \citealt{2007ASPC..376..127M}) version 5.4.0-68, Pipeline version 42030M.

\section{Data analysis}

\subsection{Overview of new results}

First, we created a dust continuum image from all data using CASA task {\tt tclean}, which resulted in a null-detection.
A new $3\sigma$ upper limit on the dust continuum is 19 $\mu$Jy beam$^{-1}$, a factor of 3 improvement over the previous limit (H18).
Next, we created a data cube from the Cycle 6 SPWs supposed to contain the [O~{\sc iii}] line using {\tt tclean} with natural weighting and 50-km s$^{-1}$ velocity binning.
We successfully confirmed the [O~{\sc iii}] line in this independent dataset.
The redshift is $z=9.1111\pm0.006$, which is consistent with $z=9.1096\pm0.006$ found by H18.
Combining all SPWs of H18 and Cycle 6 data, which contain the [O~{\sc iii}] line, we also created a data cube using {\tt tclean} with natural weighting and 50-km s$^{-1}$ velocity binning.
We use this {\it dirty} imaging data cube throughout this {\it Letter}.\footnote{
We use the {\it dirty} beam to convolve the dynamical model in \S4, avoiding any biases induced by modeling of the {\it clean} beam for the {\it clean} data cube, which is not straightforward for the combined datasets of different array configurations.}

\begin{figure*}
\centering
\includegraphics[width=7cm]{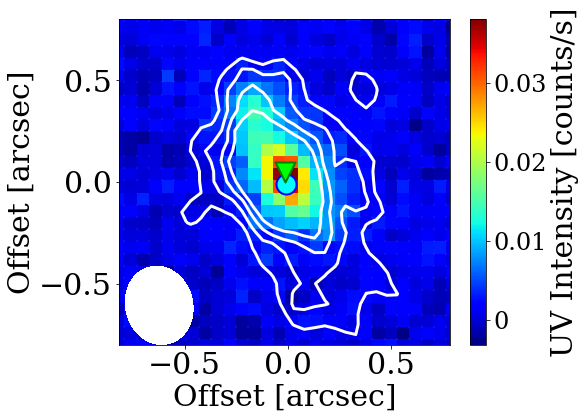}
\includegraphics[width=7cm]{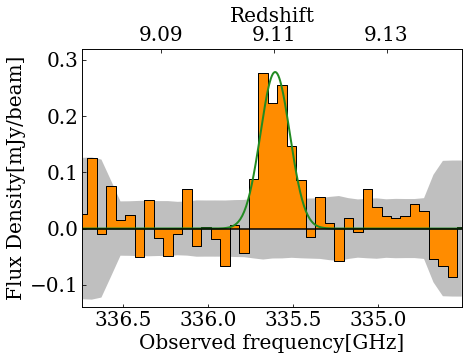}
\includegraphics[width=7cm]{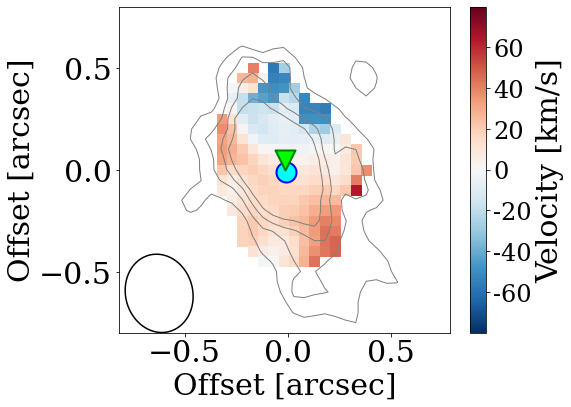}
\includegraphics[width=7cm]{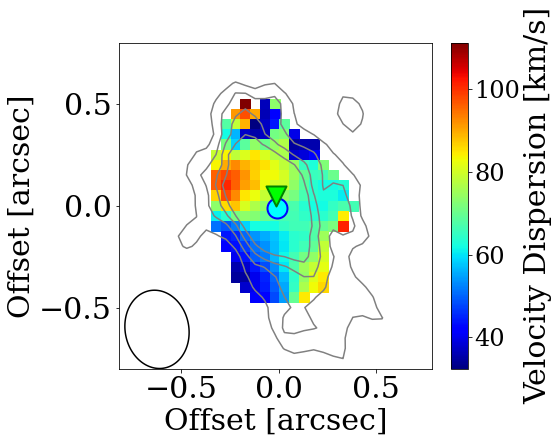}
\caption{{\it Top left}: ALMA [O~{\sc iii}] 88-$\mu$m emission moment0 map contours on the HST/WFC3 F160W image of MACS1149-JD1 at $z=9.1$. The contours show $+3, 4, 5$, and $6\sigma$ with $\sigma=9.4$ mJy km s$^{-1}$ beam$^{-1}$. The synthesized beam ellipse is shown in the bottom left corner. The circle and inverse-triangle indicate the centers of [O~{\sc iii}] emission dynamical disk (\S4) and ultraviolet (UV) emission, respectively.
{\it Top right}: [O~{\sc iii}] line spectrum integrated over the area where the line was detected at $>3\sigma$. The gray shaded region indicates the $\pm1\sigma$ noise level. The solid curve is the best-fit Gaussian profile.
{\it Bottom left}: [O~{\sc iii}] line velocity field overlaid on the line moment0 contours. The velocity field is depicted only in the area where the Gaussian line profile fitting is favored with confidence $>5\sigma$ (see \S3).
{\it Bottom right}: [O~{\sc iii}] line velocity dispersion map overlaid on the line moment0 contours. The depicted area is the same as the velocity field.
\label{fig:fig1}}
\end{figure*}

We created a velocity-integrated intensity map, i.e. moment0 map, of the [O~{\sc iii}] line (Figure~\ref{fig:fig1}, top left) using CASA task {\tt immoments} from the data cube within a velocity range of $-150$ to $+200$ km~s$^{-1}$ relative to the line redshift $z=9.1096$ (H18).
We also extracted the total line spectrum (Figure~\ref{fig:fig1}, top right) from the data cube integrated over the area where the line emission was detected at $>3\sigma$ in the moment0 map.
The velocity dispersion of the total line profile was measured at $\sigma_{\rm tot}=72.7\pm8.1$ km~s$^{-1}$, also consistent with $65.4\pm16.6$ km~s$^{-1}$ in H18.

In the moment0 map, the peak signal-to-noise ratio increased from $7.4\sigma$ (H18) to $10\sigma$, and the beam full width at the half maximum (FWHM) improved from $0.''62\times0.''52$ (H18) to $0.''39\times0.''33$.
The deconvolved FWHM of the emission was measured at $0.''81\times0.''47$, consistent with H18.
Therefore, the ionized gas is distributed smoothly without clumps, even in the improved resolution.

\subsection{Velocity field}

To examine the velocity structure of MACS1149-JD1, we adopted a method of  \citet{2018Natur.553..178S}: a Gaussian line profile fit for the spectrum at each spatial pixel in the dirty data cube.
We selected spatial pixels, where a $>5\sigma$ significance of the Gaussian fit was obtained.
Namely, $\Delta\chi^2=\chi^2_{\rm no~line}-\chi^2_{\rm Gauss}\geq25$, where $\chi^2_{\rm no~line}=\sum_{i}\frac{|I_{\rm obs}(V_i)-I_{\rm cont}|^2}{{\delta_{I_{\rm obs}(V_i)}}^2}$ and $\chi^2_{\rm Gauss}=\sum_{i}\frac{|I_{\rm obs}(V_i)-\{I_{\rm cont}+I_{\rm Gauss}(V_i)\}|^2}{{\delta_{I_{\rm obs}(V_i)}}^2}$. 
$I_{\rm obs}(V_i)$ and $I_{\rm Gauss}(V_i)$ are, respectively, the observed intensity and Gaussian line profile at velocity $V_i$. 
$I_{\rm cont}$ is the continuum level, and we set it to zero because of its null-detection. 
$\delta_{I_{\rm obs}(V_i)}$ is the observed uncertainty of the intensity at velocity $V_i$, which was measured as the root-mean-square (RMS) in the velocity channel of the dirty cube.
The resultant maps of the line velocity and the velocity dispersion are shown in the bottom left and right panels in Figure~\ref{fig:fig1}, respectively.
There is a clear velocity gradient along the north--south direction.
The maximum velocity difference was measured at $\Delta V_{\rm obs}=122\pm30$ km s$^{-1}$.
The uncertainty was calculated from those of the reddest and bluest velocities.
The velocity dispersion ranges from a few tens to a hundred km s$^{-1}$, and its average is $\sim70$ km s$^{-1}$.

\begin{figure}
\centering
\includegraphics[width=8cm]{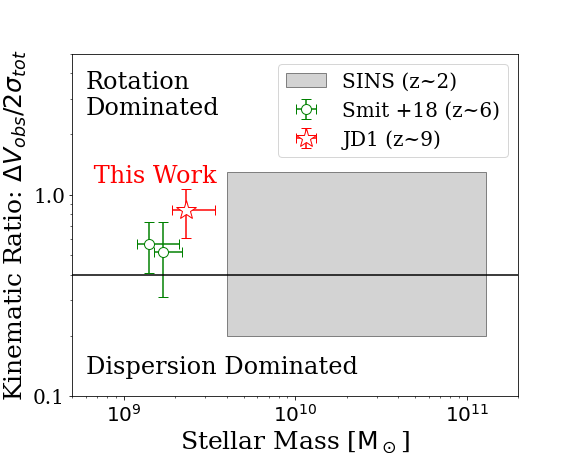}
\includegraphics[width=8cm]{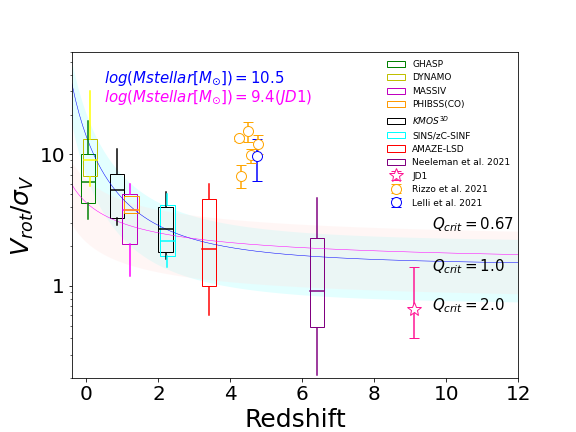}
\caption{
(Top) Observed kinematic ratio, $\Delta V_{\rm obs}/2\sigma_{\rm tot}$, as a function of the stellar mass. Our measurement of MACS1149-JD1 shown by the five-pointed star is compared with those of two $z\sim6$ galaxies \citep{2018Natur.553..178S} shown by the circles and the distribution of measurements of $z\sim2$ star-forming galaxies \citep{2009ApJ...706.1364F} shown by the shaded area. The horizontal line at $\Delta V_{\rm obs}/\sigma_{\rm tot}=0.4$ is an empirical boundary between the rotation-dominated and dispersion-dominated systems \citep{2009ApJ...706.1364F}.
(Bottom) Kinematic ratio, $V_{\rm rot}/\sigma_V$, based on dynamical modeling (\S4) as a function of redshift. The estimation for MACS1149-JD1 is shown by the five-pointed star. The box plots at $z<4$ are taken from \cite{2015ApJ...799..209W} and show the median (middle bar) as well as the central 50-percentile (box) and 90-percentile (vertical lines) of the sample distribution in each survey (see \citealt{2015ApJ...799..209W} for the references). The box plot at $z\sim6$ is the result of quasar host galaxies \citep{2021ApJ...911..141N}. The circles at $z\sim4.5$ are results of massive rotational galaxies \citep{2021Sci...371..713L,2021MNRAS.507.3952R}. The blue and magenta solid lines show semi-empirical models based on Toomre's disk instability parameter, $Q_{\rm crit}=1.0$ (quasistable thin gas disk) \citep{2015ApJ...799..209W} with a stellar mass of $10^{10.5}$ M$_\odot$ and $10^{9.4}$ M$_\odot$, respectively. The shaded areas around the lines indicate the range of $Q_{\rm crit}=0.67$ (thick gas disk) and $Q_{\rm crit}=2.0$ (star and gas composite disk) cases \citep{2015ApJ...799..209W}.
\label{fig:kinematicratio}}
\end{figure}

We have obtained a kinematic ratio of $\Delta V_{\rm obs}/2\sigma_{\rm tot}=0.84\pm0.23$, which is a factor of 1.5 greater than those observed in two rotation-dominated galaxies at $z\sim7$ reported by \citet{2018Natur.553..178S}, as shown in the top panel of Figure~\ref{fig:kinematicratio}.
The ratio is also $\sim2\sigma$ above a criterion for determining whether a galaxy is rotation- or dispersion-dominated, $(\Delta V_{\rm obs}/2\sigma_{\rm tot})_{\rm crit}=0.4$, empirically derived from a set of simulations and H$\alpha$ line observations of galaxies (e.g., \citealt{2009ApJ...706.1364F,2018Natur.553..178S}).

\section{Dynamical modeling}

\subsection{Procedure}

Motivated by the large kinematic ratio indicative of a rotation-dominated system (\S3), we performed dynamical modeling of MACS1149-JD1, assuming a geometrically thin rotating disk.
The model constraint was obtained by fitting the 3D dirty data cube, rather than the two-dimensional (2D) velocity field, to avoid rotational velocity underestimation and velocity dispersion overestimation due to the beam smearing effect \citep{2015MNRAS.451.3021D}.
The fitting procedure comprises four steps: 
(1) construction of a model of the 3D emission line data cube in the source plane, 
(2) coordinate mapping from the source plane to the image plane using a gravitational lensing model,
(3) convolution with the dirty beam profile in the image plane, and (4) optimization.
Before explaining each step below, we summarize the 9 fitting parameters in the modeling: 
$M_{\rm disk}$ (mass), 
$r_{\rm disk}$ (scale-length), 
$i_{\rm disk}$ (inclination), 
$PA_{\rm disk}$ (position angle), 
$x_{0},~y_{0}$ (central position) 
$A_{0}$ ([O~{\sc iii}] intensity at the disk center), 
$\sigma_V$ (velocity dispersion), and 
$\Delta V_{\rm sys}$ (velocity offset from the systemic redshift).

The construction of the line data cube was made by adopting Freeman's formula for the velocity field along the radial coordinate \citep{1970ApJ...160..811F}.
Namely, we assumed a geometrically thin, self-gravitating rotation disk with an exponential surface mass distribution along the radial distance, which is described by $M_{\rm disk}$, $r_{\rm disk}$, $i_{\rm disk}$, and $PA_{\rm disk}$.
The maximum rotation velocity of the disk, $V_{\rm rot}$, is calculated from $M_{\rm disk}$ and $r_{\rm disk}$ by $V_{\rm rot}=0.88\sqrt{GM_{\rm disk}/2r_{\rm disk}}$ \citep{1970ApJ...160..811F}, where $G$ is the gravitational constant.
We allowed a spatial offset of the disk center, $(x_{0},y_{0})$, from a reference point in the source plane.
The [O~{\sc iii}] line velocity field was given by this disk model, and the line profile was assumed to be a Gaussian function with a constant velocity dispersion, $\sigma_V$, throughout the system.
We allowed a constant velocity offset, $\Delta V_{\rm sys}$, from the line redshift of $z=9.1096$ (H18).
The velocity-integrated [O~{\sc iii}] line intensity distribution is assumed to follow the surface mass distribution of the dynamical disk with the last parameter describing the central line intensity, $A_0$.

Coordinate mapping was performed by adopting lensing models (convergence, $\kappa$, and shear, $\gamma_1$ and $\gamma_2$) of the MACS1149 cluster released by the Hubble Frontier Field (HFF) project \citep{2017ApJ...837...97L}.\footnote{\url{https://archive.stsci.e.,du/pub/hlsp/frontier/macs1149/models/}}
We confirmed that all six lensing models in the HFF gave qualitatively same results.
In this {\it Letter}, we present the results with the model of \cite{2015ApJ...804..103K} based on {\tt glafic} \citep{2010PASJ...62.1017O}.

Convolution with the dirty beam in the image plane was performed to the constructed line data cube model.
We adopted a synthesized dirty beam model at a velocity of 0 km~s$^{-1}$ in the observed dirty data cube.

Optimization was performed using a least-square method based on the Levenberg--Marquardt algorithm in a package of {\tt scipy.optimize.least{\_}squares}.\footnote{\url{https://docs.scipy.org/doc/scipy/reference/generated/scipy.optimize.least_squares.html}}
The model fitting was done in the image plane by comparing the modeled dirty cube with the observed cube in the velocity range of $-150$ to $+200$ km s$^{-1}$ with a 50-km s$^{-1}$ binning.
The spatial area used in the fitting was the region where the [O~{\sc iii}] line was detected at $>3\sigma$ (i.e. the region enclosed by the outer-most contour in the top left panel of Figure~\ref{fig:fig1}).
The chi-square was defined by $\sum_{i,j,k} |I^{\rm obs}_{i,j}(V_k)-I^{\rm model}_{i,j}(V_k)|^2/{\delta_{I^{\rm obs}_{i,j}(V_k)}}^2$, where $I^{\rm model}_{i,j}(V_k)$, $I^{\rm obs}_{i,j}(V_k)$, and $\delta_{I^{\rm obs}_{i,j}(V_k)}$ are, respectively, the model intensity, the observed intensity, and the observed RMS at the spatial pixel $i,j$ and the velocity $V_k$.

The initial parameter set is crucial to obtain a converged solution.
We adopted an iterative method to ensure convergence, repeating the 3D fitting where the best-fit parameters in the previous cycle were injected as the initial parameter set.
The initial guess for the first cycle was obtained from a preparatory 2D fitting for the line moment0 map and velocity field.
The best-fit parameters and their uncertainties converged within eight cycles.

Uncertainties of the fitting parameters were obtained by a Monte Carlo method.
We repeated 3D fittings for mock observed data cubes with randomly-produced image-plane noise maps.
The noise was spatially correlated on the appropriate beam-scale and its RMS was scaled to the observed one.
We adopted the best-fit values obtained from the real observed data cube as the most likely solution and the values in the central 68-percentile in the distribution obtained by the Monte Carlo method as their uncertainties.

In addition, we separately performed an exponential model fitting to the UV continuum image of the Hubble Space Telescope (HST) Wide Field Camera 3 (WFC3)/F160W filter.
The fitting parameters are the central brightness, $A_{\rm UV}$, scale-length, $r_{\rm UV}$, inclination, $i_{\rm UV}$, and position angle, $PA_{\rm UV}$.
We also allowed a spatial offset of the UV central position $(x_{\rm UV,0},y_{\rm UV,0})$ from the reference point in the source plane.
A Gaussian point-spread-function (PSF) with an FWHM of $0.''15$ of the HST/WFC3 IR channel \citep{2014ApJS..214...24S} was adopted for the convolution in the fitting procedure in the image plane.
In this particular case, we did not perform a Monte Carlo estimation of the uncertainties.

\subsection{Result}

\begin{figure*}
\centering
\includegraphics[width=18cm]{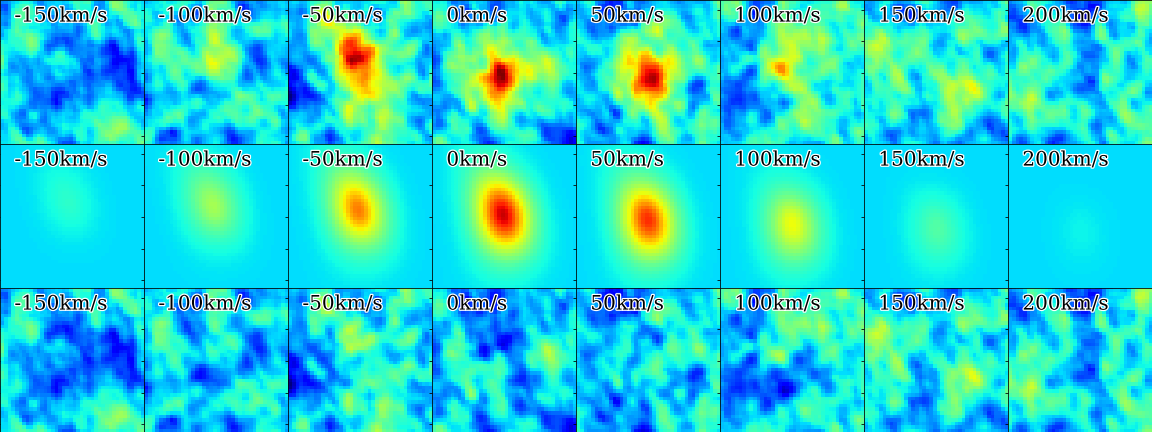}
\caption{Image plane channelmap from $-150$ km s$^{-1}$ (left) to $+200$ km s$^{-1}$ (right) with a 50-km s$^{-1}$ step. {\it Top}: the observed cube; {\it middle}: the model cube; {\it bottom}: the residual.}
\label{fig:channelmap}
\end{figure*}

\begin{figure*}
\centering
\includegraphics[width=18cm]{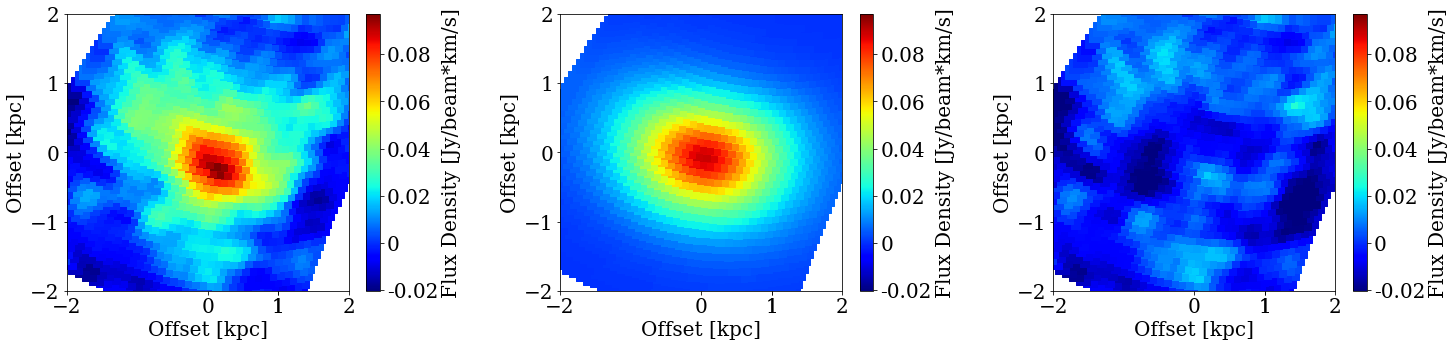}
\includegraphics[width=18cm]{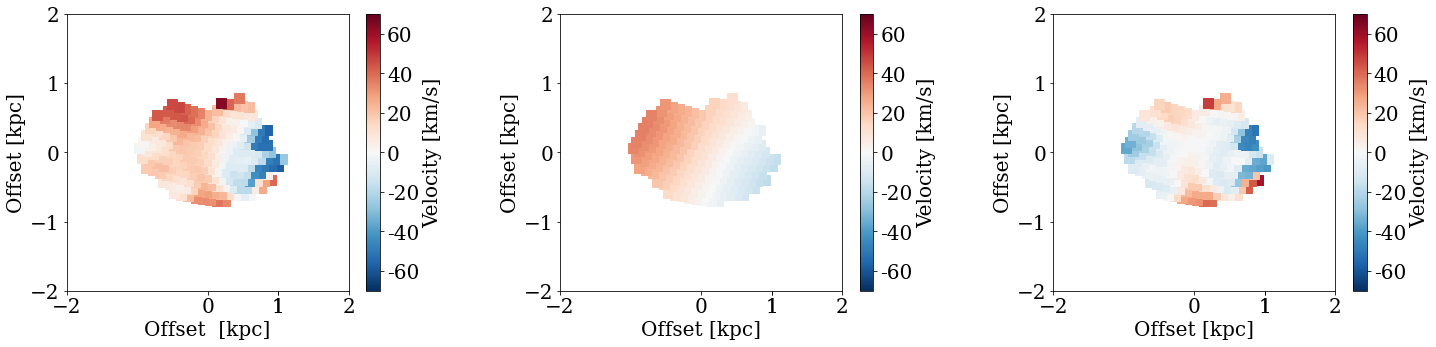}
\includegraphics[width=18cm]{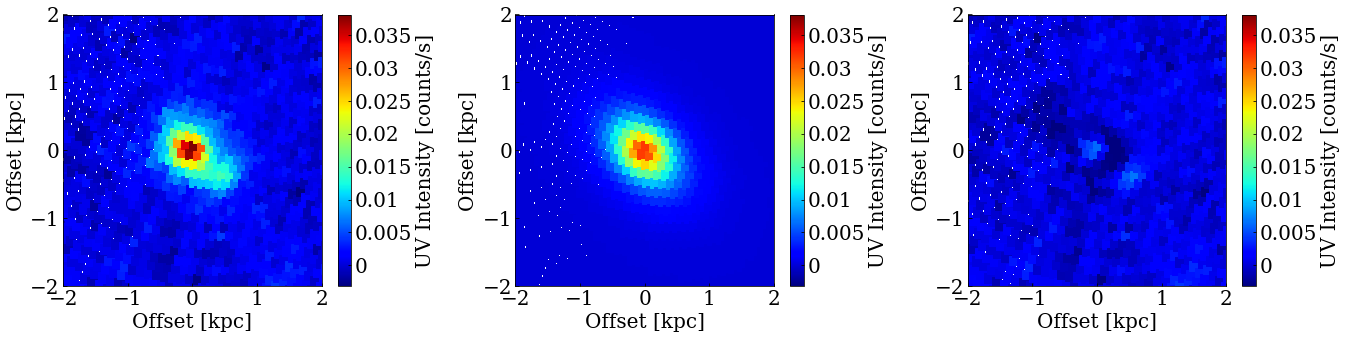}
\caption{{\it Top row}: the [O~{\sc iii}] 88 $\mu$m line moment0 map, {\it middle row}: the [O~{\sc iii}] line velocity field, and {\it bottom row}: the UV continuum image. {\it Left}: the observations, {\it middle}: the models, and {\it right}: the residuals. All images are shown in the source plane corrected for the gravitational lensing effect.}
\label{fig:SourcePlaneReconstruction}
\end{figure*}

Figure~\ref{fig:channelmap} shows a comparison between the observed and best-fit [O~{\sc iii}] line cubes in the image plane.
Figure~\ref{fig:SourcePlaneReconstruction} shows the source-plane reconstruction of the 2D [O~{\sc iii}] line moment0 map and velocity field from the best-fit line cube as well as the UV intensity map.
Low residuals shown in these figures demonstrate a reasonably good fit.
Table~\ref{tab:properties} presents a summary of the observed and derived properties of MACS1149-JD1.
The inclinations, PAs, and central positions of the [O~{\sc iii}] disk and the UV disk coincide with each other, while the scale-lengths are different.

We have found $V_{\rm rot}/\sigma_V=0.67_{-0.26}^{+0.73}$, whose range still permits a value greater than 1 consistent with a rotation-dominated system.
This ratio is compared with similarly estimated values at lower redshift from the literature in the bottom panel of Figure~\ref{fig:kinematicratio}. 
$V_{\rm rot}/\sigma_V$ of star-forming galaxies at $z<4$ and $z\sim6$ quasar host galaxies are well-explained by a semi-empirical model with a range of Toomre's disk instability parameter of $0.67<Q_{\rm crit}<2$ \citep{2015ApJ...799..209W}.
An extrapolation of the model indicates that MACS1149-JD1 at $z=9.1$ has a quasi-stable disk composed of stars and gas: $Q_{\rm crit}\sim2$ \citep{2015ApJ...799..209W}.

\begin{deluxetable}{ccl}
\tablenum{1}
\tablecaption{A summary of properties of MACS1149-JD1.\label{tab:properties}}
\tablewidth{0pt}
\tablehead{
\colhead{Property} & \colhead{Value} & \colhead{Remark}
}
\decimalcolnumbers
\startdata
\multicolumn{3}{c}{Observed values}\\
R.A. (ICRS) & $11^h 49^m 33.58s$ & H18 \\
Dec. (ICRS) & $+22^\circ 24' 45.7''$ & H18 \\
$z_{\rm spec}$ & $9.1096\pm0.0006$ & H18 \\
$M_*$ [$10^9$ M$_\odot$] & $1.08^{+0.53}_{-0.18} \times (10/\mu)$ & H18 \\
$SFR$ [M$_\odot$ yr$^{-1}$] & $4.2^{+0.8}_{-1.1} \times (10/\mu)$ & H18 \\
$\Delta V_{\rm obs}/2\sigma_{\rm tot}$ & $0.84\pm0.23$ & This work\\
\hline
\multicolumn{3}{c}{Assumed value: Lensing magnification}\\
$\mu$ & 4.7 & {\tt glafic} model \\
\hline
\multicolumn{3}{c}{Best-fit values and their 68-percentile of the dynamical model.}\\
$\Delta V_{\rm sys}$ [km s$^{-1}$] & $10.3$ ($-0.5$--$20.2$) & \\
$V_{\rm rot}$ [km s$^{-1}$] & $46.7$ ($28.3$--$87.1$) & ($\dagger$) \\
$\sigma_V$ [km s$^{-1}$] & $69.9$ ($55.6$--$75.9$) & \\
$V_{\rm rot}/\sigma_V$ & $0.67$ ($0.41$--$1.40$) & ($\ddagger$) \\
$M_{\rm disk}$ [$10^9$ M$_\odot$] & $0.65$ ($0.25$--$2.02$) & (s)\\
$r_{\rm disk}$ [kpc] & $0.50$ ($0.40$--$0.61$) & (s)\\
$i_{\rm disk}$ [$^\circ$] & $56.8$ ($44.9$--$66.3$) & \\
$PA$(disk) [$^\circ$] & $157.5$ ($143.2$--$168.4$) & \\
Reduced $\chi^2$ (Data cube) & 1.18 & \\
\hline
\multicolumn{3}{c}{Best-fit values of the UV emission model.}\\
$r_{\rm UV}$ [kpc] & $0.31$ & (s)\\
$i_{\rm UV}$ [$^\circ$] & $52$ & \\
$PA$(UV) [$^\circ$] & $138$ & \\
$\Delta$(UV--disk) [kpc] & $0.13\pm0.2$ & (s) (*)\\
Reduced $\chi^2$ (UV) & 1.08 & \\
\enddata
\tablecomments{H18: \cite{2018Natur.557..392H}. {\tt glafic} model: \cite{2015ApJ...804..103K,2010PASJ...62.1017O}. ($\dagger$): Derived from $M_{\rm disk}$ and $r_{\rm disk}$, not a fitting parameter. ($\ddagger$): Derived from $V_{\rm rot}$ and $\sigma_V$, not a fitting parameter. (s): Values in the source plane. (*) Systematic uncertainty in astrometory between the HST UV image and ALMA data cube is $0.''1$ (H18), corresponding to $\sim0.2$ kpc in the source plane.}
\end{deluxetable}

\section{Discussion}

MACS1149-JD1 satisfies an empirical criterion for a rotation-dominated system, $\Delta V_{\rm obs}/2\sigma_{\rm tot}>0.4$ \citep{2009ApJ...706.1364F}.
Among the five criteria for a rotation disk \citep{2015ApJ...799..209W}, the galaxy also satisfies (i) a continuous velocity gradient along a single axis, (ii) $V_{\rm rot}/\sigma_V>1$ (but marginally), (iv) an agreement between photometric (UV) and kinematic ([O~{\sc iii}]) axes ($<30^\circ$), and (v) a positional agreement of the dynamical center ([O~{\sc iii}]) and continuum centroid (UV).
However, it does not satisfy criterion (iii) a positional agreement between the dynamical center and the velocity dispersion peak.

For the important criterion (ii), our dynamical model gives $V_{\rm rot}/\sigma_V=0.67$ ($0.41$--$1.40$, 68\% range), still indicative of a rotation, taking into account the possible range.
We also examined a case where the [O~{\sc iii}] emission and the dynamical disk are two different components in the light of the two-component stellar populations discussed in H18.
Namely, we considered 5 additional parameters of the central positions ($x,y$), inclination, PA, and scale-length for the [O~{\sc iii}] emission disk ``decoupled'' from the dynamical disk introduced in \S4.1.
As a result, we obtained a higher kinematic ratio of $V_{\rm rot}/\sigma_V=1.4$.
The dynamical disk center was spatially offset from the centers of [O~{\sc iii}] emission and UV continuum arisen by the star-forming population, as found in cosmological simulations (e.g., \citealt{2018MNRAS.481L..84M}).
Although the fitting uncertainty was large due to the larger number of free-parameters, this decoupled disk scenario would be very interesting to examine with {\it James Webb Space Telescope} (JWST).

The criterion (iii) should also be discussed.
The velocity dispersion peak is located $\sim1$ kpc away from the [O~{\sc iii}] disk center, possibly indicating a merger or an outflow.
A possible small kink in UV emission around the dispersion peak (Fig.~\ref{fig:fig1}) might be a sign of a merger.
The weaker Ly$\alpha$ line showing blueshift from the [O~{\sc iii}] line (H18) may be another sign of a different component in the galaxy.
However, there is no distinct structure in [O~{\sc iii}] emission.
We could not examine this possibility further with the current dataset. 
The higher spatial resolution and sensitivity offered by JWST is required to address the merger possibility.
In the following, we consider that MACS1149-JD1 is a rotating disk galaxy.

The stellar mass of MACS1149-JD1 corrected by the {\tt glafic} model ($\mu=4.7$) is $2.3^{+1.1}_{-0.4}\times10^9$ M$_\odot$.
The bulk of the mass is associated with the $\sim300$ Myr-old mature stellar population, whereas $\sim3$ Myr-old star-forming population is a minor component (H18).
Crucially, the dynamical disk mass of $0.65_{-0.40}^{+1.37}\times10^9$ M$_\odot$ is consistent with that independently-determined for the mature stellar population.
If we consider the effect of the velocity dispersion in the case of $V_{\rm rot}/\sigma_V\sim1$ (e.g., \citealt{2010ApJ...725.2324B}), the dynamical mass can be $\sim5$ times larger,\footnote{From equation~(11) in \cite{2016ApJ...826..214B}, one can derive the dynamical mass corrected for the turbulent pressure as $M_{\rm dyn}=M_{\rm disk}\times\{1+4.4(\sigma_V/V_{\rm rot})^2\}$ for $M_{\rm disk}$ and $V_{\rm rot}$ evaluated at $r=2.2r_{\rm disk}$ where the Freeman disk has the maximum rotation velocity \citep{1970ApJ...160..811F}.} yielding $\sim3\times10^9$ M$_\odot$, that is also consistent with the stellar mass of $\sim2\times10^9$ M$_\odot$ (H18).
Therefore, we conclude that the dynamical mass is attributable to the mature stellar population that formed at $z\sim15$.

A cold gas component should exist in the galaxy because of the ongoing star formation.
The gas mass fraction can be expressed as $f_{\rm gas}=(a/Q_{\rm crit})(\sigma_V/V_{\rm rot})$ with $a=1$--2, depending on the velocity radial profile \citep{2011ApJ...733..101G,2015ApJ...799..209W}.
Because $Q_{\rm crit}=1$--2 for a stellar-plus-gas disk (e.g., \citealt{2007ApJ...660.1232K,2011ApJ...733..101G}), the obtained $V_{\rm rot}/\sigma_V$ suggests $f_{\rm gas}\gtrsim0.3$.
Considering uncertainties in the estimated masses, it is possible that $\sim1\times10^9$ M$_\odot$ cold gas component and $\sim2\times10^9$ M$_\odot$ mature stellar population coexist, yielding $f_{\rm gas}\sim0.3$ in a total mass of $\sim3\times10^9$ M$_\odot$.
The star formation rate of the galaxy is $8.9^{+1.7}_{-2.3}$ M$_\odot$ yr$^{-1}$ for the magnification $\mu=4.7$ (H18).
Hence, we obtain a gas depletion time of $t_{\rm dep}\sim100$ Myr when the gas mass of $1\times10^9$ M$_\odot$, which agrees with an extrapolation of a mean relation of $t_{\rm dep}\sim1.5/(1+z)$ Gyr \citep{2015ApJ...799..209W} to $z=9$.

The scale-length in the source plane of [O~{\sc iii}] disk is 1.6-times larger than that of UV continuum.
Notably, both scale-lengths should be readily resolved by the ALMA beam and HST PSF sizes in the source plane ($\sim0.3$ and $\sim0.1$ kpc in radius, respectively).
The [O~{\sc iii}] emitting ionized gas should be powered by the young stellar population traced by UV.
The extended distribution of the ionized gas compared to the young star cluster possibly suggests a significant escape of ionizing photons to a larger-scale, and the galaxy may contribute to cosmic reionization.
This may also explain the blueshift of the Ly$\alpha$ line (H18).

We have considered only a single mass component in the dynamical disk modeling.
Because we are considering the very central part ($r\lesssim1$ kpc) of the galaxy, the dark matter contribution is generally negligible (e.g., \citealt{1985ApJ...295..305V}).
According to \cite{2019MNRAS.488.3143B}, the dark matter halo mass for a stellar mass of $\sim10^9$ M$_\odot$ galaxy at $z=9$ is $\sim10^{11}$ M$_\odot$.
The virial radius of such a halo would be $\sim7$ kpc in the proper coordinate \citep{2018PhR...780....1D}.
Therefore, our observations do not reach the halo scale yet; hence, neglecting its contribution is reasonable.

In conclusion, MACS1149-JD1 at $z=9.1$, is the most distant galaxy with a signature of rotation.
This is not contradictory to the concordance cosmological structure formation.
Some theoretical studies predicted such a rotational disk in the earliest universe (e.g., \citealt{2006ApJ...645..986R,2019MNRAS.487.5902K}).
It is interesting to understand the role of two stellar populations proposed by H18 in the rotation disk formation.
We suggest that the mass of the rotational disk is dominated by the mature stellar population formed in the first major star formation episode in this galaxy at $z\sim15$ (H18).
JWST's Guaranteed Time Observation programs targeting this galaxy will resolve the different spatial distributions of the young and mature stellar populations and confirm (or revise) the scenario.

\begin{acknowledgements}
We thank Renske Smit for a discussion about the method for examining the velocity structure, Masamune Oguri for a discussion about gravitational lensing models, Yuxing Zhong for a discussion about {\tt 3D Baroro}.
AKI, YS, and YF are supported by NAOJ ALMA Scientific Research Grant Numbers 2020-16B.
TH was supported by Leading Initiative for Excellent Young Researchers, MEXT, Japan (HJH02007) and by JSPS KAKENHI Grant Number (20K22358 and 22H01258)
RSE acknowledges funding from the European Research Council under the European Union Horizon 2020 research and innovation programme (grant agreement No. 669253).
NL acknowledges the Kavli foundation.

This paper makes use of the following ALMA data: ADS/JAO.ALMA\#2015.1.00428.S and ADS/JAO.ALMA\#2018.1.00616.S. ALMA is a partnership of ESO (representing its member states), NSF (USA) and NINS (Japan), together with NRC (Canada), MOST and ASIAA (Taiwan), and KASI (Republic of Korea), in cooperation with the Republic of Chile. The Joint ALMA Observatory is operated by ESO, AUI/NRAO and NAOJ.
\end{acknowledgements}

\bibliography{reference}{}
\bibliographystyle{aasjournal}



\end{document}